\documentstyle[epsfig,12pt,a4p]{article}

\newcommand{\pip}{\mbox{$\pi^-p$ }}
\newcommand{\pt}{\mbox{$P_T$ }}
\begin{document}
\begin{titlepage}
\def\footnoterule{\hrule width 1.0\columnwidth}
\begin{center} {\large EUROPEAN ORGANIZATION FOR NUCLEAR RESEARCH}
\end{center}
\begin{tabbing}
put this on the right hand corner using tabbing so it looks
 and neat and in \= \kill
\> {Revised version} \\
\> {June 4, 1997}
\end{tabbing}
\bigskip
\bigskip
\begin{center}{ {\bf CONFIRMATION OF A SOFT  PHOTON
SIGNAL IN EXCESS OF Q.E.D. EXPECTATIONS IN \pip
INTERACTIONS AT 280  GeV/c}
}\end{center}
\bigskip
\begin{center}{
A.\thinspace Belogianni$^{  1}$,
W.\thinspace Beusch$^{   4}$,
T.J.\thinspace Brodbeck$^{   7}$,
D.\thinspace Evans$^{  3}$,
B.R.\thinspace French$^{  4}$,
A.\thinspace Jacholkowski$^{   4}$,
J.B.\thinspace Kinson$^{   3}$,
A.\thinspace Kirk$^{   3}$,
V.\thinspace Lenti$^{   2}$,
R.A.\thinspace Loconsole$^{   2}$,
V.\thinspace Manzari$^{   2}$,
I.\thinspace Minashvili$^{   6}$,
V.\thinspace Perepelitsa$^{   5}$,
N.\thinspace Russakovich$^{   6}$,
P.\thinspace Sonderegger$^{   4}$,
M.\thinspace Spyropoulou-Stassinaki$^{   1}$,
G.\thinspace Tchlatchidze$^{   6}$,
G.\thinspace Vassiliadis$^{   1}$,
I.\thinspace Vichou$^{   1}$,
O.\thinspace Villalobos-Baillie$^{   3}$
}\end{center}\bigskip
\bigskip
\begin{tabbing}
aba \=   \kill
1 \> \small
Athens University, Physics Department, Athens, Greece. \\
2 \> \small
Dipartimento di Fisica dell'Universit\a`{a} and Sezione INFN, Bari,
Italy.\\
3 \> \small
University of Birmingham, Physics Department, Birmingham, U.K.\\
4 \> \small
CERN, European Organization for Nuclear Research, Geneva,
Switzerland. \\
5 \> \small
ITEP, Moscow, Russia. \\
6 \> \small
JINR, Dubna, Russia. \\
7 \> \small
School of Physics and Chemistry, University of Lancaster, Lancaster, U.K.\\
\end{tabbing}

\bigskip
\bigskip\begin{center}{\bf {{\bf Abstract}}}\end{center}
Photons produced in \pip
interactions  at 280 GeV/$c$  were detected by reconstructing
the $e^+e^-$  pairs produced via the materialisation of the photons
in a 1 mm thick lead sheet  placed in front
of the MWPC's  of  the OMEGA spectrometer at CERN.
A soft photon signal $ 7.8\pm 1.5$ times the Q.E.D. inner  bremsstrahlung
prediction was observed confirming the results of a previous
experiment.
\bigskip
\bigskip\begin{center}{{Submitted  to Phys. Lett. B.}}
\end{center}
\end{titlepage}
\setcounter{page}{2}
\newpage
\section{Introduction}

The first observation of a soft photon  signal in excess of the Q.E.D. inner
bremsstrahlung prediction  was reported  in a $K^+p$  hydrogen bubble
chamber  experiment at 70 GeV/c\cite{ref1}.
Since then apparently  conflicting results  have been published,
some experiments observe
a signal of soft photons\cite{ref2,ref3} whilst
others observe no  excess
signal\cite{ref4}.
It appears that these results can be  reconciled if the signal
exists only in the region $Y_{cms} > 0$.
Experiment WA83\cite{ref3}, which was performed in order to confirm  the
original $K^+p$ BEBC result, found a signal of one  soft photon per
six \pip interactions at 280 GeV/c  in the kinematic region
$0.2 < E_{\gamma} < 1$ GeV/${c}$ and \pt$ < 10 $  MeV/$c$.
This is a factor 7.9 times the Q.E.D. inner bremsstrahlung prediction.
\par
In experiment WA83 the
soft photons were detected with a Pb-scintillating fiber calorimeter.
The origin of the photons was assumed to be the  main
interaction vertex for the determination of their otherwise unmeasurable
direction. The rapidity range covered was $ 1.4 \leq  Y_{cms} \leq 5$.
The calculated probability of background soft photon production
from various sources in
the apparatus was  found to be  negligible in this rapidity range.
However, we think that a direct measurement of the photon production point
would give an experimental check of this type of background.
The present experiment, using a different technique, but exploring the
same kinematic region, was motivated by this idea.
\par
    Thus, we present in this paper the results of the study of the
soft photon production in the energy range of $ 0.2 < E_{\gamma} < 1$ GeV,
and in the angular (polar angle $\theta $) range restricted to 20 mrad.
The conversion of photons to $e^+ e^-$ pairs in a thin lead sheet was
used to detect the photons.

\noindent
\section{Experimental technique}

    The data for this experiment were  collected using the set-up of the
WA91  experiment  in the CERN OMEGA spectrometer  shown in figure 1.
A 280 GeV/c {$\pi^-$} beam  incident on a 60 cm long hydrogen target,
was used to reproduce similar conditions to the WA83 exposure.
The magnetic field ( B = 1.1 T ) direction was along the z  (vertical) axis
of the OMEGA coordinate system
in which the beam is along the x-axis.
All interaction triggers (minimum bias) were collected.
Out of them only the interactions with  less than 8  charged tracks
have been used for analysis. This requirement was necessary in order to
select the cleaner events for which
the pattern recognition results are safe.
A $ 50\times50 $ cm$^2$ Pb sheet of 1 mm thickness
was placed
just before the B set of MWPC's at a distance of 73 cm  downstream
from the centre of the hydrogen target.
\par
The photons were detected via the materialisation in the Pb sheet
into an electron-positron pair. The $e^+  e^-$ were reconstructed
as $V^0$'s
from the digitisations produced in the MWPC's using  a modified
version of the standard TRIDENT\cite{ref5} reconstruction program  which
enabled  reconstruction of tracks originating in the Pb sheet
with momenta down to 40  MeV/c.
It was thus possible to
determine  the line of flight of the photon  with
an average error of $\pm$~10 mrad by measuring its momentum.
This error is  mainly due to the multiple scattering of the
electrons and positrons in the lead sheet in the energy range
$  0.2 < E_{\gamma} < 1$ GeV
used in the present analysis.
However, for the production angle of the soft photon projected on the
x-y plane ($\theta_{R}$), we preferred the more  precise  direction,
given by using the reconstructed interaction point in the hydrogen
target and the photon materialisation  point in the lead sheet.
This gives an error of $ \pm 1$ mrad.
Analogously, for the calculation of the photon polar angle $\theta$  and the
associated variable of transverse momentum
\pt the geometrical coordinates have been used.
\par
The  requirements for accepting  a positive and negative particle
pair ( $V^0 $ )  as  a materialised photon candidate were:
\begin{description}
\item [a)]
that each track had at least 4 reconstructed  space points;
\item [b)]
that the mass of the $V^0$ found by the TRIDENT reconstruction
program, assuming electron masses for the tracks,
was  less than  70 MeV/${c^2}$;
\item  [c)]
that the  x coordinate of the photon apex, defined to be
at the  position where the two  tracks
have zero angle between them, was within $\pm 3$ cm of the middle of the
lead sheet;
\item  [d)]
that the distance between the two tracks in the  x-y  plane
when the tracks had zero angle between them was  less than 3 mm, and
\item  [e)]
only those photons were taken where  the spatial  separation
of their $e^+e^-$  vertex  from
any charged track at the lead sheet was greater than 3 mm in the  x-y
plane. This was necessary to avoid  including soft photons which were
produced  in the lead sheet  by electrons or positrons originating
upstream of the sheet. Moreover, this cut suppressed a large
fraction of the bremsstrahlung photons radiated by these particles
upstream of the lead sheet, since both the parent particle and its
radiation arrive on the lead sheet with a small separation.
\end{description}
\par
The above defined  criteria (c), (d) and (e) have been arrived at
after a visual scanning of a large number of reconstructed
real events including  soft photon candidates
and MC simulated events (using the simulation described
in the next paragraphs).
\par
The efficiency for reconstructing photons was determined
by a method involving  implanting simulated photons into the real data .
The method  generates photons with a bremsstrahlung-like spectrum,
converts them in the lead sheet using
the EGS4 code\cite{ref6} , transports the resulting $e^+e^-$ pairs through
the lead sheet and the MWPC's  and simulates clusters in the MWPC's
at the position where the  $e^+$  and $e^-$  cross the MWPC's.
After digitizing these clusters were implanted  on  actual events
which passed through the TRIDENT reconstruction program followed by a
standard selection and analysis algorithm.
The efficiency has been studied as a
two-dimensional function of energy and emission angle
of the photons, with non-equidistant binning (down to
50 MeV in energy and 0.5 mrad in $\theta$).
The error on the overall correction factor is estimated
to be less than 20\% even in the bins with the smallest statistics.
The validity of the efficiency  correction  can be assessed   by
comparing  the efficiency  corrected photon \pt spectrum, in a region
of \pt where photons  from hadronic  decays dominate,
with the predictions of the FRITIOF Monte Carlo program for hadronic
interactions\cite{ref7} as we will see below (Sect.4).

We used this Monte Carlo technique to prove that the $P_T$
measurement, based on the $\gamma$ apex coordinate finding, is not
distorted by multiple scattering. The $P_T$ error comes mainly from the
reconstruction error in the $\gamma$ apex $z$ coordinate,
and was found to be $\pm 2.4$ MeV/$c$, with the $\theta$ accuracy of
$\pm 5.6$ mrad. As to the $\gamma $ energy measurement, it was
accurate to within 5 MeV (All errors quoted in this paragraph are
average and relevant to the photon energy range of $0.2 - 1$ GeV).

\noindent
\section{Study of backgrounds}

The following sources of soft photons were considered:
\begin{description}
\item [a)]
inner bremsstrahlung;
\item [b)]
photons from hadronic decays;
\item [c)]
Dalitz pairs from $\pi^0, \eta$ and $\omega$ decays;
\item [d)]
knock-on electrons from energetic tracks;
\item [e)]
spurious $V^0$ consisting of hadronic tracks which nevertheless satisfy
the photon selection criteria.
\item [f)]
secondary photons: when a high energy photon generates an $e^+ e^-$
pair in the material upstream of the lead sheet (target, target walls,
Silicon detector, etc.) the pair particles may radiate bremsstrahlung
photons, which can enter our kinematic region. In most cases such
photons come to the lead sheet close to their parent charged paricles
and are rejected by the isolation cut (cut (e), Sect.2).
However, there is a fraction
of such photons which are not rejected because their parent particles
bend in the OMEGA magnetic field. Additionally, pairs from
photons of $E_\gamma > 1$ GeV converted in the lead sheet can degrade
to energies below 1 GeV due to bremsstrahlung.
\end{description}

In order to calculate the yields of soft photons from (a) to (f)
we have developed a Monte Carlo program which transports the particles
generated by the FRITIOF code [7] ($\gamma$'s and Dalitz $e^+ e^-$
pairs included) through our experimental set up. The EGS4 code [6]
was involved for the proper treatment of the $e^+ e^-$ pairs and photons.
A generator of inner bremsstrahlung photons was added to the FRITIOF code.
The bremsstrahlung calculations were based on:
\begin{description}
\item [i)]
the exact Low formula [8a]
\begin{equation}
\frac{d\sigma}{d^{3}\vec{k}}
=
\frac{\alpha}{(2 \pi)^2} \frac{1}{ \omega}
\int d^3 \vec{p}_{1} . . . d^3 \vec{p}_{N}
\sum_{i,j} \eta_{i} \eta_{j}
\frac{ (P_{i} P_{j}) }{ ( P_{i} K )  ( P_{j} K )}
\frac{ d \sigma^{H}}{ d^{3} \vec{p}_{1} ... d^{3} \vec{p}_{N}}
\end{equation}

where $K$ and $\vec{k}$ denote photon four- and three-momenta,
$P$ and $\vec{p}$ are 4- and 3-momenta
of charged hadrons, $\eta=1$ for beam pion and for positive
outgoing particles, $\eta=-1$ for the target proton and negative
outgoing particles, and the sum being extended over all charged particles;
\item [ii)]
the Haissinski formula [8b], which is expected to be more stable with
respect to lost (undetected) particles (when it is applied to the real
data events). It has the same form as (1) with the scalar products of
4$-$vectors $(P_i P_j)$ being replaced by
$(\vec{p}_{i \bot} \vec{p}_{j \bot})$, where
$\vec{p}_{i \bot} = \vec{p}_i-(\vec{n} \vec{p}_i) \vec{n}$,
{}~$\vec{n}$ is the photon unit vector.
\end{description}

Both formulae give results coinciding within 10\%, so we take this number
as a systematic uncertainty for the bremsstrahlung calculations.

The results of the Monte Carlo estimations for the soft gamma yields
from the sources listed above are given in Table 1.

\noindent
\section{Experimental results}

 The $P_{T}^{2}$  spectrum of reconstructed photons
emitted inside a cone of half angle of 215 mrad around the beam
direction   and with energy  $ 0.2 < E_{\gamma} < 1 $ GeV,
uncorrected
for efficiency
is shown in figure 2. The shape  expected from  $\gamma$'s
of hadronic decays  in the same kinematic region
has been obtained using the FRITIOF Monte Carlo
and is shown by the dashed  line.
The Monte Carlo data have been normalised by a constant factor  to the
higher $P_{T}^{2}$  region of the experimental sample.
Since we know that the photon detection efficiency falls off with
decreasing $P_{T}^{2}$ the presence  of an increase in the observed
uncorrected spectrum at small $P_{T}^{2}$
(\pt $< 10 $ MeV/$c$
or $P_{T}^{2}  <  10^{-4}$  ( GeV/$c$)$^{2}$ )  is
evidence for a low  $P_{T}^{2}$   signal not originating  from
hadronic decays.
\par
Figure 3a shows the efficiency corrected  \pt spectrum upon which
are superimposed the predictions  of the FRITIOF Monte Carlo
increased by 20\% to fit
the data above a $P_T$ of 50 MeV/$c$. This factor
can be attributed to the residual systematic errors in the
$\gamma$  detection  and reconstruction procedure (15 \% as
was evaluated by varying the parameters of the efficiency finding code),
and to the FRITIOF Monte Carlo systematics (about 10 \%).
In this figure, while the data  follow  well the expected  \pt distribution
of photons coming from hadronic decays  above a   \pt   of 50 MeV/$c$
where the contribution
coming from Q.E.D. inner bremsstrahlung is small, below
\pt $<$50 MeV/$c$  an excess exists  which rises rapidly  towards
zero  \pt.
As can be seen  by comparing figures 3a and 3b,
this excess is  essentially concentrated in  angles
$ {\theta} <  20 $ mrad
\footnote{We note, that the number of photons selected under this cut
($\theta < 20 $ mrad) is not affected by the experimental accuracy in
the $\theta $ angle ($\pm 5.6 $ mrad) since the angular distribution of
the soft photons ($E_{\gamma} < 1$ GeV) is almost uniform around the
$\theta = 20$ mrad.}.
\par
The rate of  photons  to interactions in this kinematic  region is
$ (13.1 \pm 0.4 )$\%. The quoted error is statistical.
The systematic error , due to uncertainty in the efficiency, is 15\% of the
rate.
Reducing the photon rate by the sum of contributions  (b) to (f) of Table 1
we find  that, in the defined kinematic region, the signal of the soft photons
is $(9.8\pm 0.4 \pm 1.5)$\% , and the ratio  for the observed signal to the
expected Q.E.D. inner bremsstrahlung is $ 7.8\pm 1.5$ , where the main
contribution to the error comes from the uncertainties in the efficiency
correction and the inner bremsstrahlung calculation.
This value may be compared to the value  of $ 7.9\pm 1.4$  found
in the WA83 experiment for the same kinematic region.
\par
In order to see whether the low \pt photons observed in figure 3b  originate
from the interaction point we show, in figure 4a, a correlation plot of
$\theta_{R}$
against $\theta_{P} $ which is the production angle of the photons
defined by the line of flight of the photon in the
x-y plane as measured by the vector sum of the $e^+$  and $e^-$  momenta.
A  $45^0$  correlation is observed with a spread about this line of
$\pm $ 10 mrad, which is what is expected from the Monte Carlo study
and comes mainly from the multiple scattering of the $e^+ e^-$
in the lead sheet. The $\delta \theta = \theta_P - \theta_R$ distribution,
shown in figure 4b, was found to be of a Breit-Wigner shape which is expected
for a resolution function in this experiment (superposition of many
Gaussian distributions of a variable width depending on the multiple
scattering angle which varies with $E_{\gamma})$, [9]
\footnote{The conditions under which the resolution function was
obtained in [9] to be of a Breit-Wigner form, are very close to those
we have in the experiment: fast fall-off of the multiple scattering error
distribution with the energy increase since it comes from
the bremsstrahlung-like spectrum, and a sharp cut-off at small energies
due to decreasing detection efficiency.}
. The $ \Gamma $ (full width) of the Breit-Wigner was
fitted to be $(19.2 \pm 0.6)$ mrad, its position being centered at zero within
0.3 mrad and the fit $ \chi ^2$ being 85.6 per
$n.f.d = 92$. As can be seen from this
figure, the yield of the non-correlated photons to the signal is
consistent with zero, with the upper limit for it being less than
0.2 \% per event (at 99\% C.L.). This upper limit was obtained by adding
a constant term to the fit form (Breit-Wigner), increased from zero until
the total $ \chi ^2$ increases by
6.6, the 99 \% confidence level for the fit with a single parameter.

Thus, the fact that we observe a strong correlation  is
evidence that the photons originate from the point of interaction.
Furthermore, the angular precision offered by
$\theta_{R}$  and  $\theta_{P}$  angles   can be exploited
for the comparison of WA91 and WA83 experiments.
Having observed the same signal in the very forward
direction in both experiments,
with the interaction and materialisation  points  entering into the angle
measurement ( WA91),
we obtain  evidence that the soft  photons  in WA83 also originate
from the target interactions.  Otherwise, in WA91
where the material between target and photon detection is significantly
less than in WA83,
we should observe less signal.

\noindent
\section{Conclusion}

   This experiment confirms the  existence of an anomalous soft photon
signal  in the kinematic region
0.2 GeV $< E_{\gamma}< 1$ GeV
and  $\theta  < 20 $ mrad at a level $7.8\pm 1.5 $  times the
expected from Q.E.D.  inner bremsstrahlung.
\bigskip
\newpage
%

\newpage
\noindent
Table 1.\\
Calculated soft gamma yields ($ 0.2 < E_{\gamma} < 1$ GeV, $\theta < $ 20
mrad).
The quoted errors are statistical. The systematic errors for background
calculations are estimated to be 10 \%.
\begin{center}
\begin{tabular}{ l c c c }
\hline
 & Source & Number of $\gamma$ per event (\%) \\
 && corrected for conversion efficiency (factor 8) \\
\hline
$a$. & Inner bremsstrahlung & $ 1.25 \pm ~0.05 $ \\
$b$. & Hadronic $\gamma$'s & $ 1.53 \pm ~0.05 $ \\
$c$. & Dalitz pairs & $ 0.09 \pm ~0.01 $ \\
$d$. & Knock-on electrons & $ 0.12 \pm ~0.02 $ \\
$e$. & Spurious $\gamma$'s & $ (2 \pm 2) 10^{-3}  $ \\
$f$. & Secondary $\gamma$'s & $ 1.52 \pm ~0.05 $ \\
\hline
 & Sum over sources $a$ to  $f$ & $ 4.50 \pm ~0.09 $ \\
\hline
\end{tabular}
\end{center}

\newpage
\begin{figure}[1]
\begin{center}
\end{center}
\caption{Layout of the OMEGA spectrometer for the experiment }
\end{figure}
\newpage
\begin{figure}[2]
\begin{center}
\epsfig{figure=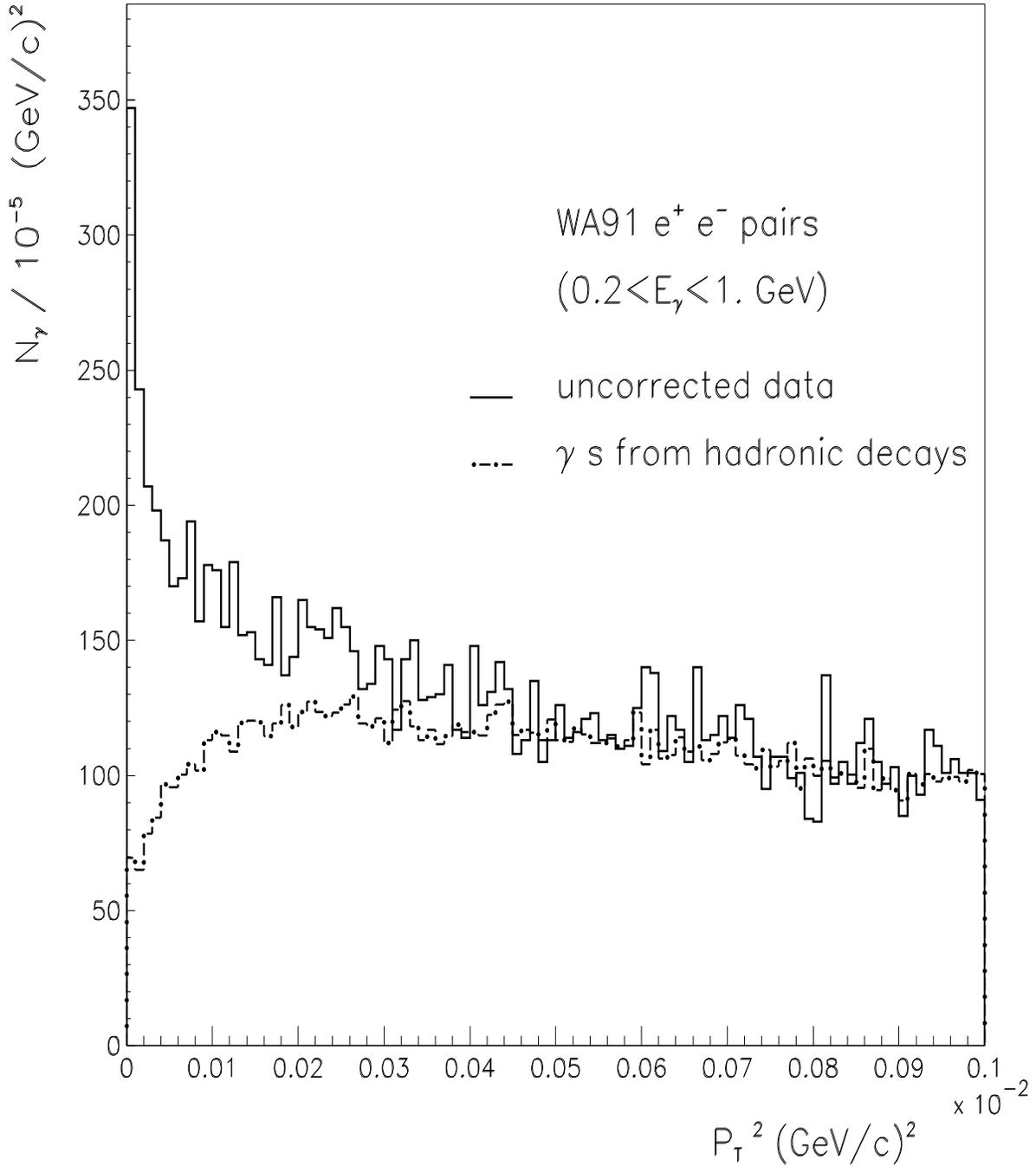,height=20cm,width=17cm}
\end{center}
\caption{$P_{T}^{2}$ distribution for photons with energy
$0.2 < E_{\gamma} <  1.0$   GeV
uncorrected  for detector efficiency. }
\end{figure}
\newpage
\begin{figure}[3]
\begin{center}
\epsfig{figure=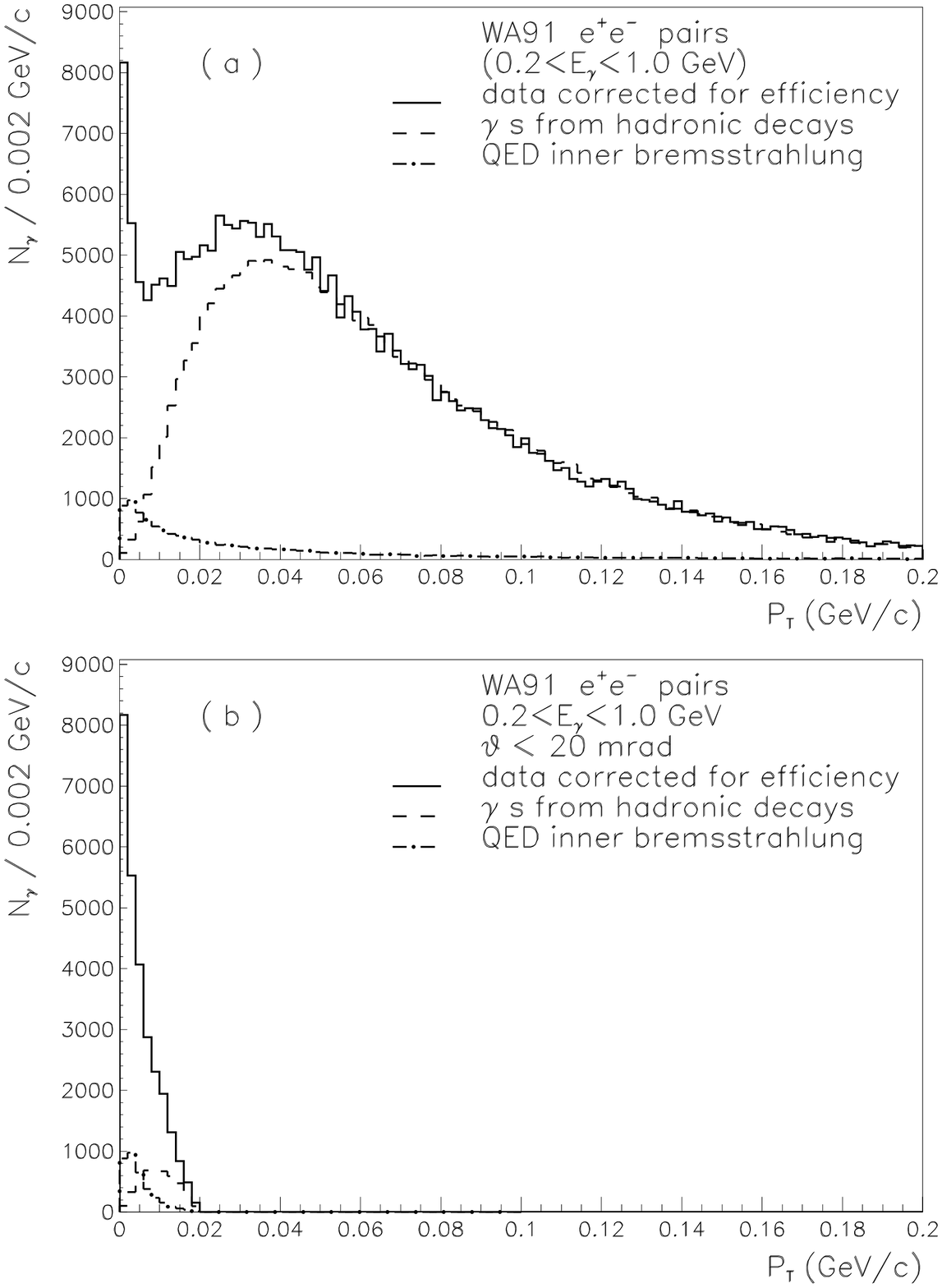 , height=23cm,width=17cm}
\end{center}
\caption{a) $P_T$  distribution for photons with energy
$0.2 < E_{\gamma} <  1.0 GeV $
corrected for detection efficiency;
b) Same as figure 3a  but with  additional  restriction of
$\theta < 20$  mrad. }
\end{figure}
\newpage
\begin{figure}[4]
\begin{center}
\epsfig{figure=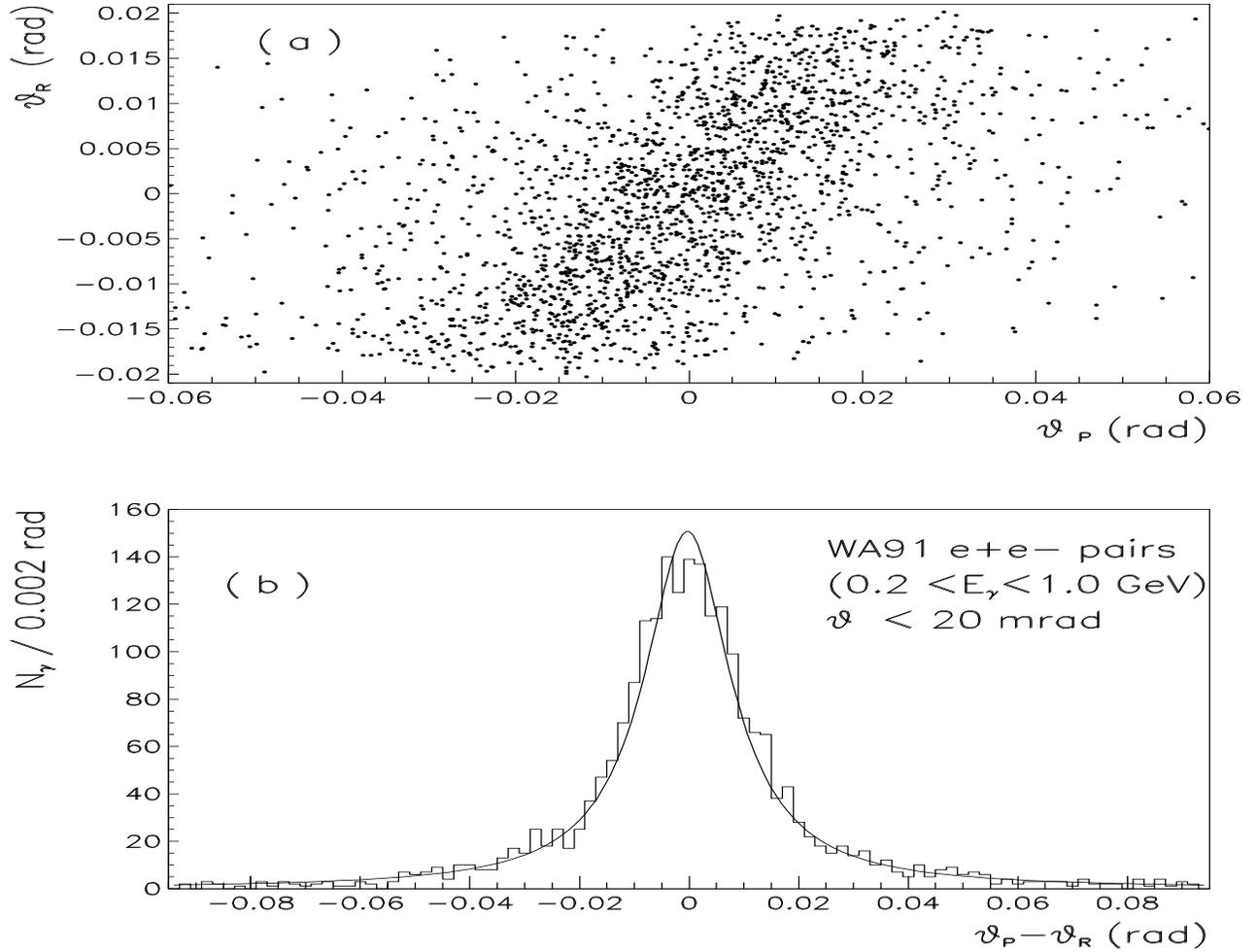   ,height=15cm,width=19cm}
\end{center}
\caption{a) Correlation of ${\theta}_{P}$  versus ${\theta}_{R}$,
as defined in the text, for the sample of $e^+e^-$  pairs in figure 3b ;
b) The difference $ \theta_P - \theta_R$ for the small $P_T$ photons
of figure 4a
(histogram) with the result of a fit with a Breit-Wigner (full line).}
\end{figure}
\end{document}